\documentclass[12pt]{article}
\usepackage{epsfig}
\usepackage{hhline}
\usepackage{times}
\usepackage{cite}

\newlength{\dinwidth}
\newlength{\dinmargin}
\begin{document}  

\begin{titlepage}
\begin{flushright}
IEKP-KA/2001-20 \\
Sept. 2001
\end{flushright}

\vspace*{1cm}

\begin{center}
\begin{LARGE}

{\bf Construction and Performance of a}

\vspace*{0.1cm}
{\bf Micro-Pattern Stereo Detector with }

\vspace*{0.1cm}
{\bf Two Gas Electron Multipliers}

\end{LARGE}
\end{center}

\vspace{0.7cm}
\noindent
\hspace*{2.4cm}
T.~Barvich,
P.~Bl\"um, 
M.~Erdmann, 
M.~Fahrer,
K.~K\"archer, \\
\hspace*{2.4cm}
F.~K\"uhn, 
D.~M\"ormann,
Th.~M\"uller, 
D.~Neuberger,
F.~R\"oderer, \\
\hspace*{2.4cm}
H.J.~Simonis, 
A.~Skiba, 
W.H.~Th\"ummel,
Th.~Weiler,
S.~Weseler$^t$

\vspace{0.5cm}

\begin{center}
{\em Institut f\"ur Experimentelle Kernphysik, 
Universit\"at Karlsruhe, \\
Wolfgang-Gaede-Str. 1, 76131~Karlsruhe, Germany}
\end{center}

\vspace*{0.5cm}

\begin{abstract}
The construction of a micro-pattern gas detector of dimensions 
$40\times 10\,$cm$^2$ is described.
Two gas electron multiplier foils (GEM) provide the internal amplification 
stages.
A two-layer readout structure was used, 
manufactured in the same technology as the GEM foils.
The strips of each layer cross at an effective crossing angle of 
$6.7\,$degrees and have a $406\,\mu$m pitch.
The performance of the detector has been evaluated in 
a muon beam at CERN using a silicon telescope as reference system.
The position resolutions of two orthogonal 
coordinates are measured to be $50\,\mu$m and 1\,mm, respectively.
The muon detection efficiency for two-dimensional space points 
reaches $96\%$.
\end{abstract}

\vspace*{0.4cm}
\noindent
\hspace*{2.4cm} Key words: detector, position sensitive, GEM, two-layer readout \\
\hspace*{2.4cm} PACS 29.40.Cs, 
     29.40.Gx 

\vspace{1.0cm}

\begin{center}
{\sf The authors would like to dedicate this work to 
Siegfried Weseler \\ 
who unexpected died of a heart attack June 2001.}
\end{center}

\end{titlepage}

\section{Introduction}

\noindent
Tracking of highly energetic particles which scatter under small 
angles in a strong axial magnetic field provides a prime challenge 
in future collider experiments.
The trajectories of these so-called ``forward particles''
in the field require excellent measurements of the azimuthal 
coordinate along the tracks to precisely determine their momenta.
For the simultaneous measurement of the radial coordinate only
moderate precision is needed to determine their scattering angles
with respect to the collision axis.  

Gas micro-pattern detectors have been developed over many 
years for such applications.
In particular, the small size readout pattern can handle 
high particle rates and provide good spatial resolution. 
Although these detectors 
were conceived for one-dimensional readout, 
two-dimensional readout has been made possible by introducing 
strips with a crossing angle or pads on the backside of the readout 
substrate \cite{angelini}. 
This method however has serious problems 
associated with the low intensity of induced signals.

A new concept of two-dimensional readout was introduced by applying 
an etched Kapton\footnote{Kapton: Polymid film (Trademark of Du Pont).} 
technology producing crossing strips on the
readout side of the substrates \cite{sauli1}.
The readout structure is kept at ground potential.
Together with gas electron multipliers (GEM), developed by 
F. Sauli \cite{sauli}, such detectors can be produced at a reasonable cost
and provide sufficient
safety margin to be operated in a high rate environment \cite{moermann}.

\section{Overview of the Detector Module }

The module which is described here is a closed system of four 
trapezoidal detector units produced on a 
common board forming a segment of a ring (called 
{\it the detector module}) with the readout electronics
and high voltage connections outside the contiguous gas volume 
of the detector. 
In between the readout structure and the drift cathode, two gas electron 
multipliers are inserted.
 
The separation of the readout stage and amplification stage 
allows the use of any appropriate readout pattern. 
While an 
orthogonal electrode system is considered as the optimum with respect 
to spatial resolution in both coordinates, it has the drawback 
of combinatorial ambiguities in case of multi-hit events. In 
order to optimize the detector for high rate capability and
good resolution in the measurements of the particle momenta and 
scattering angles, a 
small crossing angle is chosen for the readout strips. 
The geometrical dimensions of the module have been chosen to 
correspond to the outer ring 
of the CMS forward tracker \cite {mf2} in order to make as much 
use as possible of existing tools and equipment. 

\begin{figure}[htb]
\center
\setlength{\unitlength}{1cm}
\begin{picture}(15.0,7.0)
\put(2.,0.)
{\epsfig{file=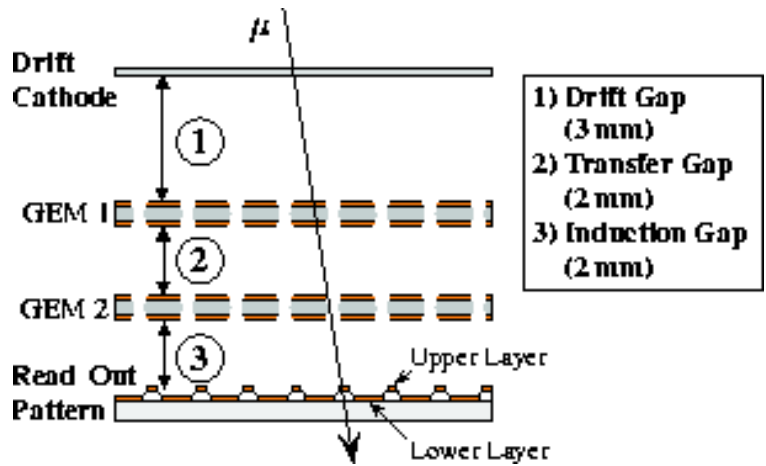,width=10cm}}
\end{picture}
\caption{Schematic view of a detector module.}
\label{fig:schema}
\end{figure}

Figure \ref{fig:schema} 
shows a schematic view of the 
detector module consisting of three gas gaps formed by frames supporting 
readout structure, GEM-foils, and drift cathode.

\subsection{The Readout Structure}
The coordinate information is extracted from the signals 
of strip electrodes produced on a 300\,$\mu$m thick 
glass fibre enforced epoxy board.
The artwork for the readout structure was produced at 
CERN\footnote{A. Gandi, R. De Oliveira, CERN-EST-SM, 
Geneva, Switzerland.  } using the same technology as for the 
production of GEM foils. In a first step, the lower 
strip layer is produced by means of a lithographic method onto a 
copper-cladded board 
to which, in a second step, a single-sided 
copper-cladded 50\,$\mu$m thick Kapton foil is glued. 
Again by lithographic means 
the strips of the upper readout layer are patterned on that foil. 
Finally, the Kapton on top of the lower strips is removed by a chemical 
etching process, and the copper strips are gold plated.

\begin{figure}[htb]
\center
\setlength{\unitlength}{1cm}
\begin{picture}(15.0,9.0)
\put(2.,0.)
{\epsfig{file=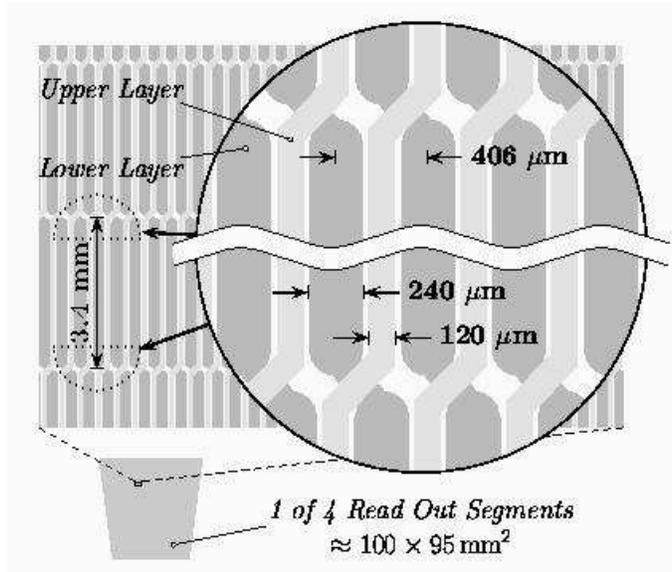,width=9cm}}
\end{picture}
\caption{Schematic view of the stereo readout structure.}
\label{fig:struct}
\end{figure}

The strips of the upper and lower readout layers
cross at a relative angle of $6.7\,$degrees. In order
to minimize the crossing area and thus the capacitive coupling 
between the two layers, the 
strips are not made as straight lines but show a ``steplike'' shape
producing up to 28 radially segmented readout pads
(see Fig.~\ref{fig:struct}). The widths of the strips are chosen to 
approximately equalize the charge sharing between the two readout 
planes resulting in 
upper strips that are 120\,$\mu$m wide and  
lower strips 240\,$\mu$m wide. The thickness of the 
copper strips is about 5\,$\mu$m. 
Each plane of the four detector units consists of 256 strips at a 
pitch of 406\,$\mu$m.

\subsection{Gas Electron Multipliers (GEM)}

The GEMs have been produced at the same CERN workshop as the readout 
structure. They consist of a 50\,$\mu$m thick Kapton foil, copper cladded 
on both sides. By lithographical methods and chemical etching,
a regular hexagonal matrix of holes has been 
produced. The pitch of the 
holes is 120\,$\mu$m and the diameter is 80\,$\mu$m in the copper 
and about 45\,$\mu$m in the Kapton. Applying a voltage 
difference between both copper sides produces a dipole 
field high enough to provide gas amplification.
Both copper sides have been etched into four electrically segmented units 
each corresponding in size to that of a readout unit. 

\subsection{Drift Cathode}

The drift cathode is made of 360\,$\mu$m thick 
Ferrozell\footnote{Ferrozell GmbH, Augsburg, Germany.} , a glass fibre 
enforced epoxy, on which
a 20\,nm thick layer of gold has been evaporated. Due to the 
low density of 1.75\,g/cm$^3$ the thickness corresponds to 250\,$\mu$m 
glass but with a much higher mechanical stability.
The thermal expansion coefficient of 
Ferrozell ($1.6 \cdot 10^{-5}\,$K$^{-1}$  ) is comparable to 
that of Stesalit\footnote{Stesalit
AG, Zullwill, Switzerland.} (the material of the support 
frames) and so no thermal stress is expected.

\subsection{Frames}

The module frames 
are manufactured of Stesalit in the mechanical workshop of the
IEKP Karls\-ruhe. 
Because of the small thickness of the drift cathode and the 
small cross section of the frames, an additional 2\,mm top frame
has been added 
to provide additional mechanical stability for supporting
the GEM foils. The spacer 
frame between drift and upper GEM has a height of 3\,mm. 
Between the two GEMs, and between the lower GEM and the readout plane,
2\,mm spacer frames are used. 
The width of the frames is 3\,mm.
While the primary ionization volume of 3\,mm thickness is 
typical to minimize the time of charge 
collection, the 2\,mm transfer and induction gaps
are optimized to separate 
the two amplification stages and to guarantee minimal 
charge diffusion. 

\section{Assembly}

The mechanical assembly has to be done thoroughly to ensure 
a reliable performance of the detector. This requires careful 
quality control of all components, strict observance of high 
grade cleanliness during assembly, 
and dedicated tools which are briefly described in the following sections.

\subsection{Quality Control}
\subsubsection{Readout Structure}

In order to identify broken strips or shorts, 
the readout structure is tested by measuring the capacitance 
between neighbouring strips within each layer, and the capacitance of pairs of 
strips from the upper and lower layer which are neighbours on the bonding 
side of the structure. 
The measurement is performed on a fully automatic probestation which 
connects the individual pairs to a precision capacitance 
meter\footnote{Keithley 590 CV Analyzer.}.
\begin{figure}[htb]
\center
\setlength{\unitlength}{1cm}
\begin{picture}(15.0,10.0)
\put(1.,0)
{\epsfig{file=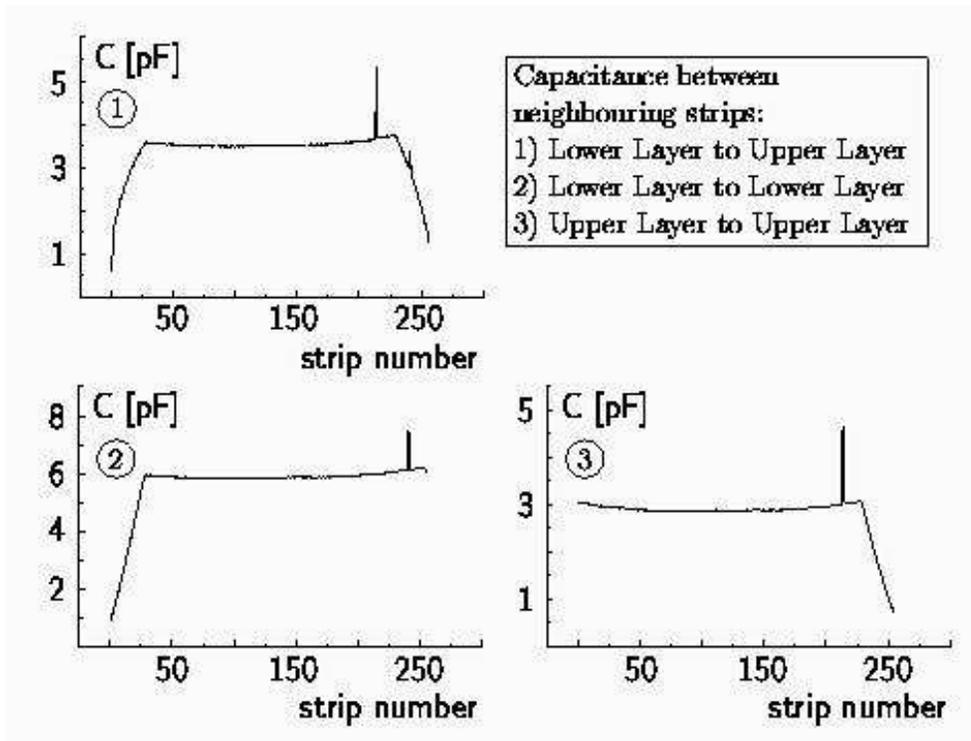,width=13.cm}}
\end{picture}
\caption{Inter--strip capacitance of an individual readout unit, showing a 
small substrate defect (see text).}
\label{fig:quality}
\end{figure}

Figure~\ref{fig:quality} shows the result of such a measurement for one 
individual readout unit. The capacitance values 
are reduced near the edges of 
the unit due to the trapezoidal shape of the readout structure 
resulting in shorter strip lengths on the non--parallel sides.
The identification of a small defective region is evident in 
Fig.~\ref{fig:quality} where the presence of a short leads to 
increased capacitance values.

\subsubsection{GEM-Foil}

The quality control of the GEMs consists of an optical inspection 
and an electrical test for shorts between both copper sides. 
Moreover, the electrical insulation is tested in a dry nitrogen 
atmosphere up to 450\,V. To reach this voltage difference a 
careful ``training'' procedure is followed. In particular 
the GEM is operated for 12 hours at 400\,V. A leakage current 
of less than 20\,nA at this setting is required for acceptance. 
As a result, 22 out of 24 delivered GEM foils were accepted.
 
\subsection{Gluing the Module}

Before mounting, all mechanical parts are cleaned in an ultrasonic bath with 
de-ionized water and dried in a laminar nitrogen flow cabinet. 
The GEM foils and the 
readout structure are thoroughly flushed with dry nitrogen immediately 
before gluing.

\begin{figure}[htb]
\centering
\epsfig{file=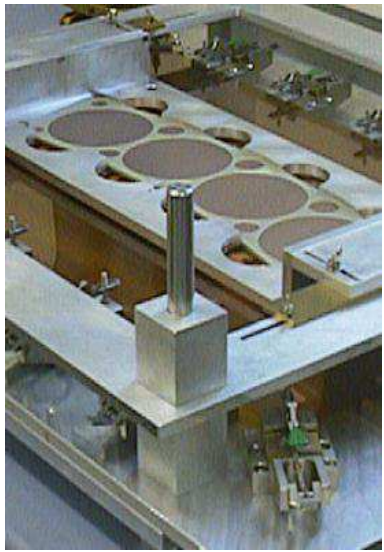,width=5cm}
\caption{View of the gluing jig. It shows the aligned top frame 
above the GEM stretched by means of spring-loaded clamps. The aligned 
spacer frame below the GEM is not visible.}
\label{fig:gemspann}
\end{figure}

Since each GEM foil is supported only by the thin spacer 
frame, particular care is given to the assembly procedure 
of the drift cathode and the GEMs in order to provide 
enough stability to keep the GEM stretched. In a first step
the drift cathode is glued simultaneously between 
top and spacer frame. After curing, this part together with 
the next spacer frame is glued to the first
GEM foil which is stretched by means of a spring-loaded 
dedicated tool (see Fig.\ref{fig:gemspann}). 

\begin{figure}[htb]
\centering
\epsfig{file=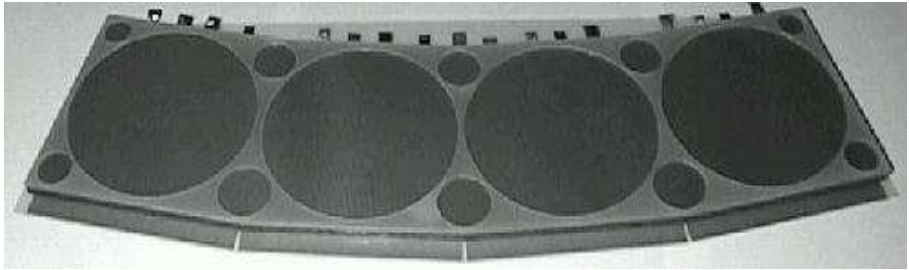,width=12cm}
\caption{The assembled detector module.
The high voltage connection pads are visible at the upper edge,
and the 4 regions each consisting of 512 readout pads are seen at the 
lower edge.}
\label{fig:kammer}
\end{figure}

After curing, this step is repeated to glue the second GEM foil.
Finally the readout board and the drift-GEM part 
are joined, closing the active volume of the detector 
module. All gluing is done using room temperature curing
EPO-Tek 302\footnote{Polytec GmbH, Waldbronn, Germany.}.
Figure \ref{fig:kammer} shows the assembled module.

\subsection{Connecting the Module}

At distinct locations, the metallization of the GEM foils and of the drift 
cathode extends to the outside of the detector gas volume, forming the 
high voltage connection pads. After soldering thin high voltage cables to 
these pads, the remaining metallized structure is passivated by covering with 
a layer of epoxy to avoid discharges.

On the low voltage side, the readout strips are wire-bonded to a pitch 
adaptor reducing the effective 203\,$\mu$m pitch from the alternating strips 
of the two readout layers to the 44\,$\mu$m pitch of the front end chips.

The readout hybrid consists of a ceramic substrate with a set of
4 PreMux front end chips \cite{premux}.
Each chip has 128 channels of charge preamplifiers, shaper-amplifiers 
and double-correlated sampling circuitry as well as an analogue multiplexer.
The peaking time of the shaper was set at $45\,$ns appropriate for high rate 
environments.
One hybrid serves each of the four individual detector units. 
Together with an additional output buffer board, a single Flash-ADC channel 
is sufficient to readout the entire detector module. 

The gas mixture is fed into the active detector volume through fan-outs
integrated in the short sides of the frames.
The gas flow was directed such that the GEM holes were flushed by the gas.

\section{Detector Performance}

\noindent
The performance of the detector was examined in a high energy muon beam at 
CERN. The aims of the measurements were to 
\begin{itemize}
\item establish the region of optimal working parameters,
\item explore the detector characteristics in view of the stereo readout 
      of the charges created in a single gas volume, and 
\item examine the tracking quality where the criteria are the position 
      resolutions, tracking efficiency and purity.
\end{itemize}

\subsection{Experimental Setup}

\noindent
The experimental setup was situated in the X5 beam line of the CERN West area. 
A low intensity muon beam of energy $100\,$GeV traversed a series of 
detectors placed on an optical bench. 
Events were triggered by a coincidence of two plastic scintillators 
of dimensions $6\times 12\,$cm$^2$ and $2\times 2\,$cm$^2$, respectively. 
A silicon telescope consisting of 4 double-sided detector layers with
50\,$\mu$m pitch, was used to measure two orthogonal coordinates \cite{bari}.
The detectors have a sensitive area of $2\times 2\,$cm$^2$ and
were positioned in two groups at a distance of 65\,cm, 
allowing for precise predictions of the muon trajectories crossing the GEM 
detector. 
The position of the GEM detector was 90\,cm behind the second
silicon group. 
The detector was operated with Ar:CO$_2$ in the ratio $70:30$ which we 
consider as a cheap gas mixture with no particular safety risks.        

\subsection{Analysis Tools and Definition of Observables}
\label{sec:anaTools}

\noindent
To analyse the responses of the detectors, the program package IRIS 
\cite{iris} was used and modified to our needs \cite{kuehn}. 
The program provides administration of the run-by-run pedestal and 
noise of each channel, and a clustering algorithm for the measured 
charge depositions.
The cluster algorithm used considers channels with ADC values 
exceeding two standard deviations of the noise level. 
Starting at the channel with the maximum charge, all contributing 
neighbouring channels are grouped into candidate clusters.
The main characteristics of the clustering algorithm are:
\begin{itemize}
\item The signal charge $q$ of each cluster is calculated from the charge 
sum of all contributing strips in units of ADC counts.
\item The cluster position is provided by the charge weighted average 
position of the contributing strips in units of strip numbers.
\item The cluster size is given by the number of consecutive strips 
contained in the cluster.
\item As a measure of the cluster noise, the noise level of the channel 
with the maximum charge in the cluster is taken.
\end{itemize}
The program treats the cluster searches in both readout layers 
of the GEM detector separately.

As a measure of the relative gain we use the most probable value $Q$
determined from fits of the Landau distribution to the data.

A further program, described in \cite{hagen}, was employed to measure the muon 
trajectories in the silicon telescope using the cluster candidates 
provided by the IRIS package.
This program was used to align the detectors and 
to predict the muon positions on the GEM detector in order to measure 
the position resolution $\sigma$ in the detector.

The program is further used to determine the tracking efficiency for 
single muons traversing the GEM detector,
\begin{equation}
\varepsilon = 
\frac{N_{GEM}(\vert x_{GEM}-x_{\mu}\vert < 5 \sigma)}{N_{\mu}} \; .
\label{eq:efficiency}
\end{equation}
Here $N_\mu$ is the number of muon tracks, and the nominator 
denotes the number of clusters in the detector which are associated 
with the predicted muon track within $5$ standard deviations of the 
measured position resolution.

The purity of the cluster measurements is determined from 
\begin{equation}
\rho = \frac{N_{GEM}(\vert x_{GEM}-x_{\mu}\vert < 5 \sigma)}
{N_{GEM}({\rm overlapping \, with \, telescope})} \; .
\label{eq:purity}
\end{equation}
In the denominator, all clusters are considered which overlap with 
the sensitive area of the silicon telescope.
 
\subsection{Signal Distributions}

\subsubsection{Charge Distributions}

\begin{figure}[htb]
\center
\setlength{\unitlength}{1cm}
\begin{picture}(15.0,14.0)
\put(1.5,0.0)
{\epsfig{file=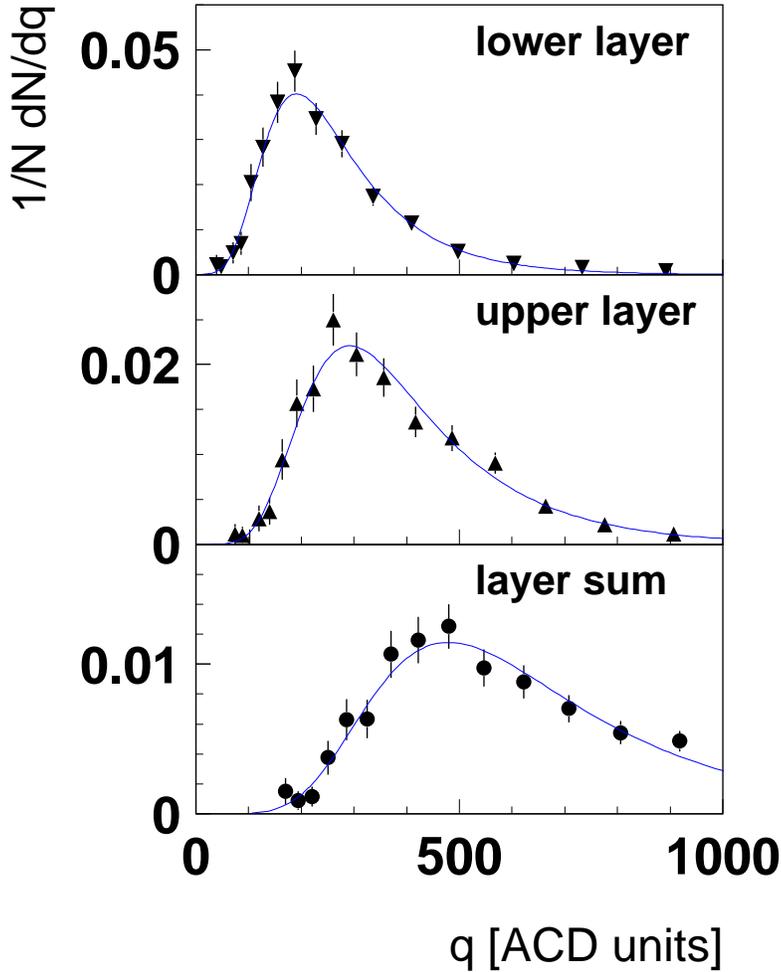,width=12.0cm}}
\end{picture}
\caption{Distributions of the cluster charge $q$ 
for the lower layer, the upper layer, and the sum of both.
The curves denote fits based on the Landau distribution.}
\label{fig:signal} 
\end{figure}

\noindent
Examples of charge distributions of the largest signal cluster in the event 
are shown in Fig.~\ref{fig:signal}. 
The down-pointing triangle symbols represent the cluster charges $q$
measured in the lower readout layer. 
The distribution is found to approximately follow a Landau distribution. 
The corresponding fit is shown by the curve. 
The signal distribution of the upper readout layer is shown by the 
up-pointing triangle symbols together with the corresponding fit. 
This layer collects more charge carriers than the lower readout layer. 
Note that by construction the pad area of the lower layer is larger than that 
of the upper layer by a factor of $2$, partially balancing the charge 
collection in the two layers. 
For this comparison, the small regions where the two readout layers cross 
have been excluded from the distributions. 
The circle symbols represent the event-by-event sum of the charges collected 
in both layers, and the curve gives the corresponding fit. 

\begin{figure}[htb]
\center
\setlength{\unitlength}{1cm}
\begin{picture}(15.0,10.0)
\put( 2.9,8){\rm\bf\large a)}
\put( 9  ,8){\rm\bf\large b)}
\put(0.,0.0)
{\epsfig{file=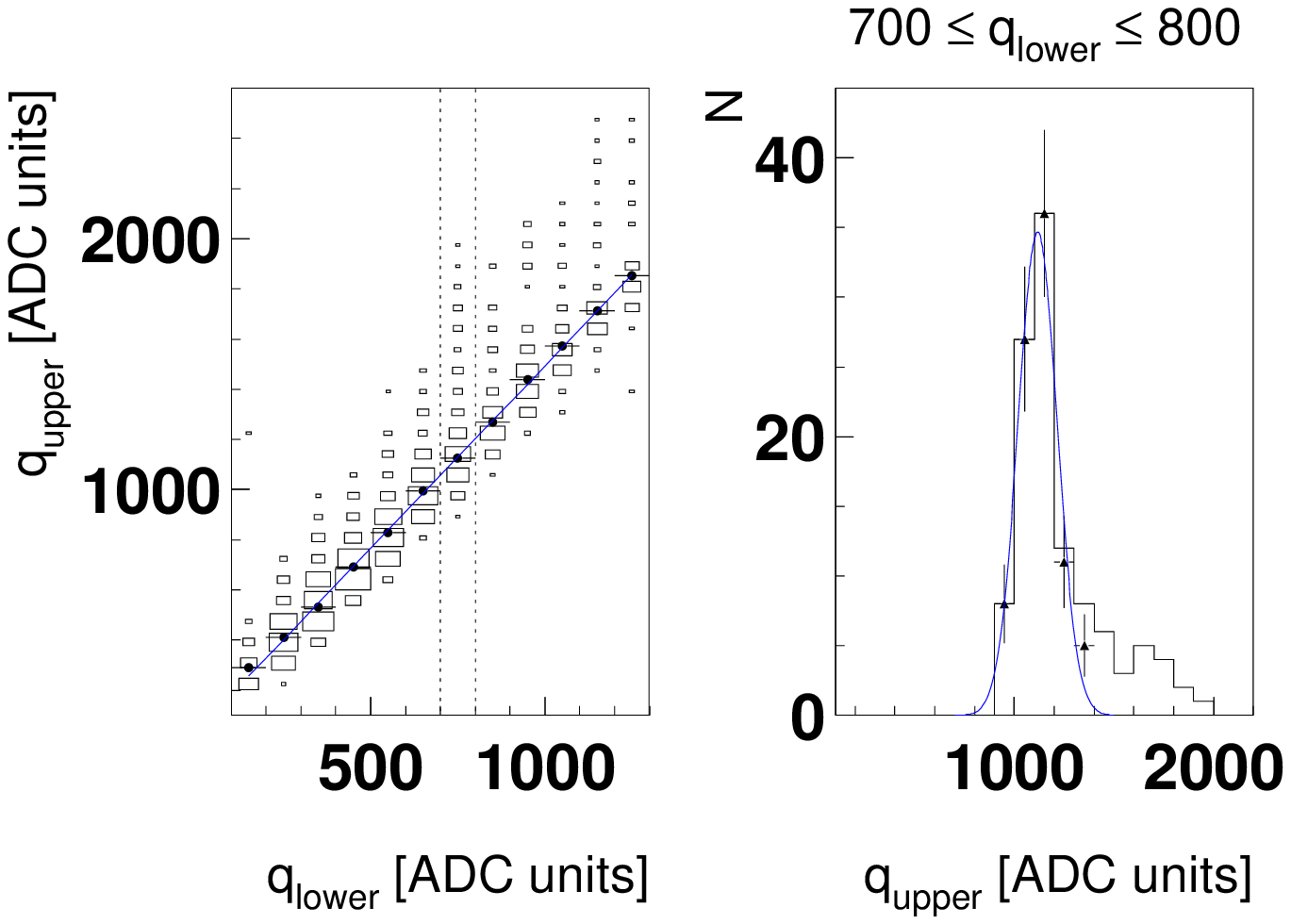,width=15.0cm}}
\end{picture}
\caption{
a) Relation of the signal charges of the upper and lower readout layers
   for the two clusters with the largest ADC values in the event.
b) Cluster charge distribution of the upper layer for the charge interval 
   $700\le q\le 800\,$ADC units of the lower readout layer (histogram).
   The symbols represent charge detection on the pads,
   excluding crossing regions of the readout layers.
   The curve shows a Gaussian fit to the latter measurement. }
\label{fig:qcorr} 
\end{figure}

The charge relation of the two readout layers is further analysed in 
Fig.~\ref{fig:qcorr}a.
The box symbols represent the largest signal clusters found in the 
upper and lower readout layers.
The distribution shows a strong correlation between the two signals together 
with a tail towards larger signals in the upper layer. 

The histogram in Fig.~\ref{fig:qcorr}b shows the charges collected in the 
upper readout layer while restricting the lower layer signals to the interval 
$700\le q \le 800\,$ADC units. 
The symbols show the charge measurements 
relating to events where the charge collection appeared within 
the region of the $3.4\,$mm long pads.
The tail towards larger signals in the upper readout layer therefore
originates from the small regions where the strips of the two layers cross. 

For charge collection within the pads, the signals of the upper 
readout layer are larger by a factor of $1.5\pm 0.1$ compared to the 
lower readout layer as is shown in Fig.~\ref{fig:qcorr}a by the  
symbols together with a linear fit to these values.
The Gaussian fit of Fig.~\ref{fig:qcorr}b gives a width of 
$\sigma_q \sim 10\%$ which demonstrates a strong charge 
correlation between the two layers. 

In the region where the two layers cross, the signal of the upper layer 
is found to be a factor of $2$ above that of the lower layer 
(not separately shown in the figure).

\subsubsection{Number of Strips Contributing to the Clusters
\label{sec:strips}}

\begin{figure}[htb]
\center
\setlength{\unitlength}{1cm}
\begin{picture}(15.0,10.0)
\put( 2.8,7.8){\rm\bf\large a)}
\put( 8.2  ,7.8){\rm\bf\large b)}
\put(0.,0.0)
{\epsfig{file=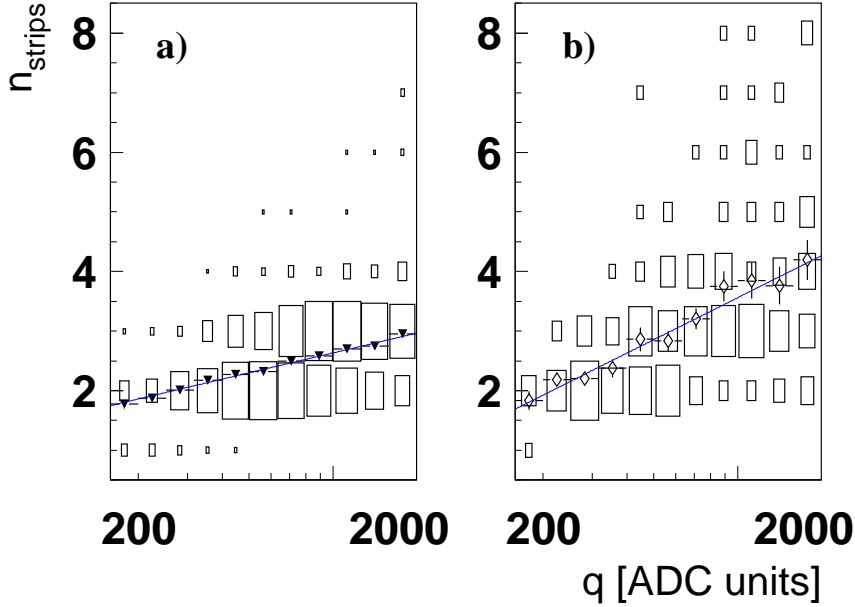,width=14.0cm}}
\end{picture}
\caption{
The box symbols show the number of strips contained in a signal
cluster of the lower readout layer as a function of the cluster charge.
The symbols give the average number of strips, and the 
curves are fits to these values.
In a) the drift field strength was above $E_D=2\,$kV/cm, 
and in b) below this value.}
\label{fig:strips} 
\end{figure}

\noindent
The spatial extent of the charge cloud collected on the readout layers 
gives additional information on the signal characteristics and the 
stability of the operating conditions.
Examples of the number of strips contributing to the cluster with the 
largest signal in the event are shown in Fig.~\ref{fig:strips} as a 
function of the cluster charge $q$.
In Fig.~\ref{fig:strips}a, the drift field was above $E_D=2\,$kV/cm.
The average cluster size (plotted as triangle symbols) 
varies between $2-3$ strips per layer and 
increases with $q$, as expected from an 
increasing number of neighbouring channels exceeding the noise cut
applied in our cluster finding algorithm.
The dependence is compatible with a 
logarithmic increase of the mean cluster size with its signal charge
and the curve shows the corresponding fit.

Below $E_D=2\,$kV/cm (Fig.~\ref{fig:strips}b), the number of strips 
contributing to the clusters are found to be much larger on average 
compared to the measurements above this drift field value.
Since the same time delay relative to the trigger signal was used in
all measurements, this effect presumably results from the electronics 
when the electron arrival times change with smaller drift velocities.
In all following studies we therefore use the distribution of the
cluster width as a criterion to ensure stable operating conditions.

\subsection{Determination of the Working Parameters}

\noindent
In order to determine the optimal field strengths and to derive the 
corresponding signal in comparison to the noise level, field scans 
have been performed and analysed with respect to the cluster with the
largest signal in each layer using the following conditions,
\begin{enumerate}
\item The scans were analysed for detector regions with a uniform noise level 
of the contributing strips of about 15 ADC counts.
This allows our cluster signal over single strip noise values 
to be converted to other definitions of the cluster noise.
\item Events where the charges are collected in the crossing region of 
the two readout layers are excluded.
The charges of the small fraction of signals from the crossing region can 
be derived with the information given above.
\item To reject clusters not associated with muons traversing the detector, 
both readout layers have to show a signal cluster in the region 
covered by the beam.
\end{enumerate}
In the case of the drift field scan, where the statistics 
have been relatively small owing to the $6\times 12\,$cm$^2$ 
scintillator used for triggering, 
only condition 1) was applied together with the requirement of a hit in 
one of the silicon layers in front of the GEM detector.

\subsubsection{Scans of the Charge Collecting Fields}

\begin{figure}[htb]
\center
\setlength{\unitlength}{1cm}
\begin{picture}(16.0,12.0)
\put(3.0,0.)
{\epsfig{file=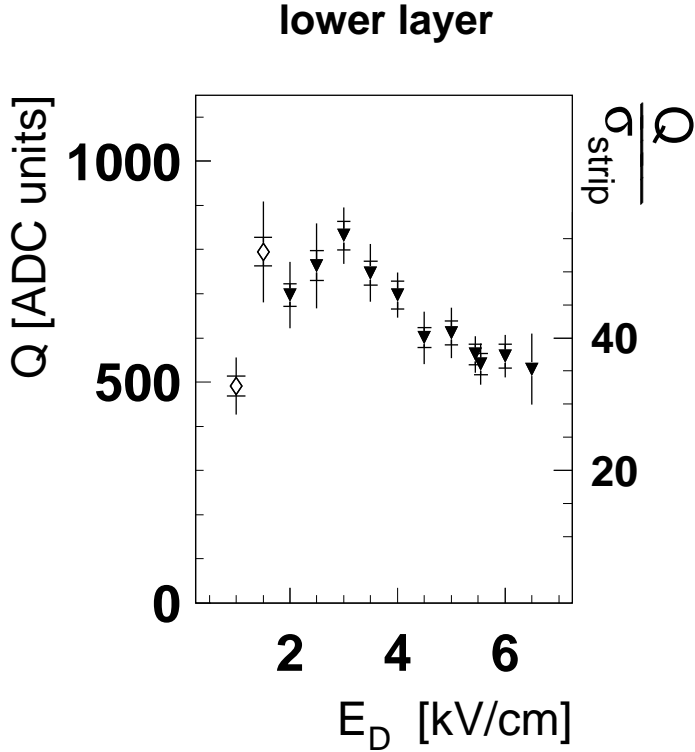,width=10.0cm}}
\end{picture}
\caption{
   Relative gain $Q$ as a function of the {\em drift field} 
   strength. The right axis indicates the signal over single strip noise 
   ratio. }
\label{fig:drift} 
\end{figure}

\noindent
{\em a) Drift Field.}
In Fig.~\ref{fig:drift}, the drift field strength has been varied 
while fixing the 
strength of the other fields 
($E_T=4.5\,$kV$/$cm, $E_I=4.3\,$kV$/$cm, $\Delta U_{GEM}=430\,$V). 
The vertical axis represents the relative gain $Q$, 
measured in the lower readout layer. 
The inner error bars represent the statistical errors of the fits, the 
total errors include a conservative estimate of the uncertainty resulting from 
varying the binning and the region of the fits. 
The scale on the right axis gives the ratio of the relative gain 
divided by the average noise of a single channel contributing to the cluster. 

The measurement exhibits a maximal $Q$ at about $E_D\sim 2-4\,$kV$/$cm. 
Here the signal over single strip noise is approximately $50$.
For values above $E_D=4\,$kV$/$cm, the signal is reduced. 
Simulations have shown that at high values of the drift field the proportion 
of the field lines ending on the upper GEM foil increases, leading to 
losses of primary electrons \cite{zittel}.

Instructive information on the region of low drift field values comes from
Fig.~\ref{fig:strips} which was described in section \ref{sec:strips}.
The open diamond symbols in Fig.~\ref{fig:drift} denote measurements
with significantly larger cluster sizes compared to the other settings
of the drift field.
Although the relative gain is already large at $E_D=1.5\,$kV/cm,
stable operating conditions start at $E_D=2\,$kV/cm.

\begin{figure}[htb]
\center
\setlength{\unitlength}{1cm}
\begin{picture}(16.0,12.0)
\put(3.0,0.)
{\epsfig{file=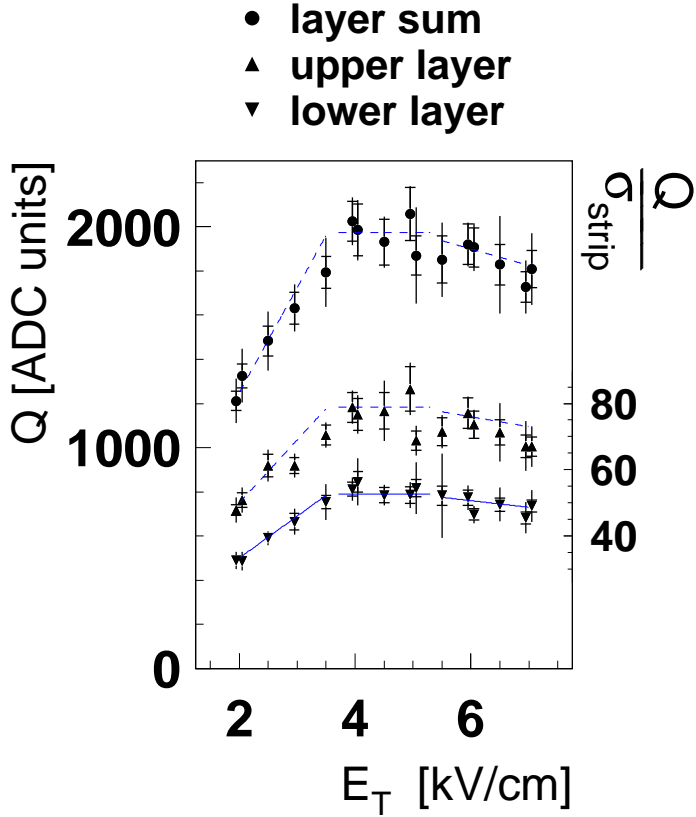,width=10.0cm}}
\end{picture}
\caption{
   The relative gain $Q$ is shown as a function of the {\em transfer field}
   strength for the lower/upper layers (down/up pointing triangle
   symbols) and the charge sum of the two layers (circles).
   The curves serve to guide the eye.
   The right axis indicates the signal over single strip noise 
   ratio.}
\label{fig:transfer} 
\end{figure}

\noindent
{\em b) Transfer Field.}
In Fig.~\ref{fig:transfer}, the relative gain $Q$ is shown as a 
function of the transfer field, keeping the other fields fixed 
($E_D=3\,$kV$/$cm, $E_I=5\,$kV$/$cm, $\Delta U_{GEM}=430\,$V). 
In this figure the down/up pointing triangle symbols represent the 
signals measured in the lower/upper readout layers. 
The circles give the charge sum of the two layers. 
The solid line gives a guidance to the eye on the measurements of the 
lower readout layer. 
The transfer field leads to maximum charge values in the region of 
$E_T\sim 4-6\,$kV$/$cm.
The corresponding signal over single strip noise is around $50$ 
in the lower readout layer and $80$ in the upper layer respectively.

The dashed lines give the expected signals in the upper layer 
and the layer sum from the charge correlation discussed in 
Fig.~\ref{fig:qcorr}. 
The good agreement of the upper layer measurements with this 
prediction shows that the charge correlation is a characteristic of 
the detector and is independent of the field values explored here. 

The average cluster width was found to increase with the cluster 
charge $q$ at a similar rate as seen in Fig.~\ref{fig:strips}a,
implying good operating conditions for all measurements of the transfer 
field.

\begin{figure}[htb]
\center
\setlength{\unitlength}{1cm}
\begin{picture}(16.0,12.0)
\put(3.0,0.)
{\epsfig{file=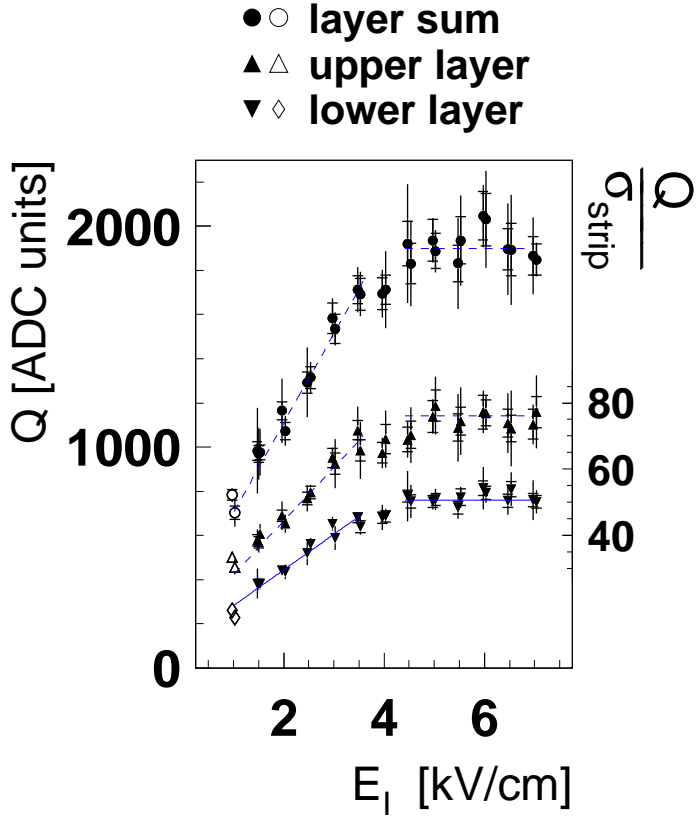,width=10.0cm}}
\end{picture}
\caption{
   The relative gain $Q$ is shown as a function of the {\em induction field}
   strength for the lower/upper layers (down/up pointing triangle
   symbols) and the charge sum of the two layers (circles).
   The curves serve to guide the eye.
   The right axis indicates the signal over single strip noise ratio.}
\label{fig:induction} 
\end{figure}

\noindent
{\em c) Induction Field.}
In Fig.\ref{fig:induction}, the dependence of the relative gain $Q$ 
is shown as a function of the induction field strength $E_I$. 
The other fields remained fixed at
$E_D=3.5\,$kV$/$cm, $E_T=4.5\,$kV$/$cm, and $\Delta U_{GEM}=430\,$V.
The signals increase with increasing $E_I$, reaching a plateau for 
$E_I>4\,$kV$/$cm, with a signal over single strip noise ratio of about 
$50$ and $80$ in the lower and upper readout layers respectively. 
The measurements remain in this plateau, consistent with the field lines 
ending on the readout structure. 

An analysis of the cluster widths showed much larger clusters, on average,
at the lowest induction field value (open symbols) compared to all
other measurements (full symbols).
Good operating conditions were reached for $E_I\ge 1.5\,$kV/cm.

\subsubsection{Scans of the Gas Electron Multiplier Fields}

\begin{figure}[htb]
\center
\setlength{\unitlength}{1cm}
\begin{picture}(16.0,12.0)
\put(3.0,0.)
{\epsfig{file=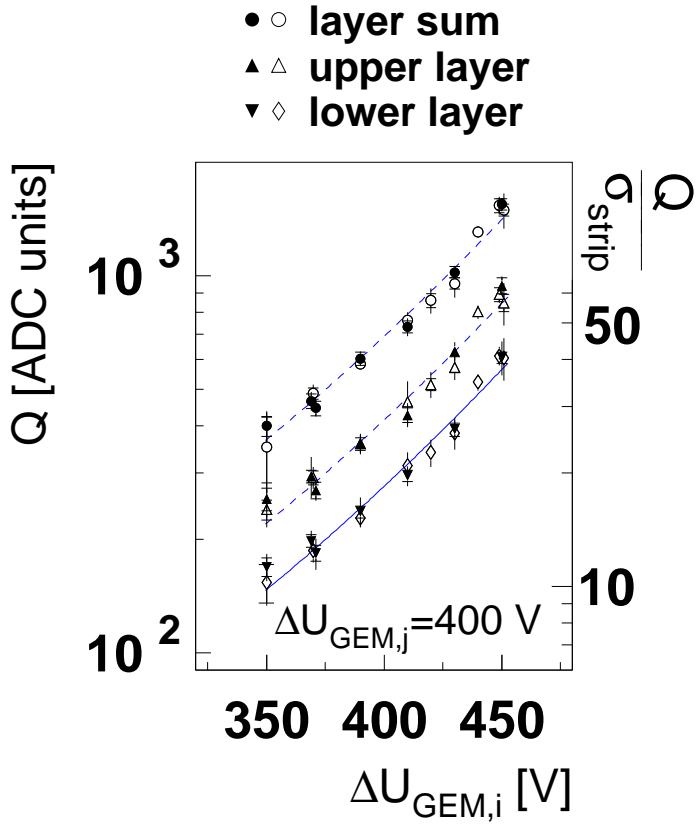,width=10.0cm}}
\end{picture}
\caption{
   The relative gain $Q$ is shown as a function of the voltage
   applied to the GEM foils for the lower/upper 
   layers (down/up pointing triangle symbols) and the charge sum of the 
   two layers (circles).
   The closed/open symbols denote the variations in the first/second foil.
   The right axis indicates the signal over single strip noise ratio.
   For the curves refer to the text.}
\label{fig:gemscan} 
\end{figure}

\noindent
In Fig.~\ref{fig:gemscan}, the dependence of the relative gain $Q$ on the 
voltage applied to the two sides of the first GEM foil is shown by the 
solid symbols. 
The other fields were at
$E_D=3.5\,$kV$/$cm, $E_T=4.5\,$kV$/$cm, $E_I=5.5\,$kV$/$cm, and
$\Delta U_{GEM2}=400\,$V.
The open symbols show the corresponding scan of the second GEM voltage, 
keeping $\Delta U_{GEM1}=400\,$V. 
The measurements exhibit a strong increase in dependence of the applied
voltage as expected for gas electron amplification
and the behaviour is the same when varying the first or second GEM voltage. 
To guide the eye, 
the solid curve represents an exponential fit to the lower readout layer
measurements which was chosen to be 
quadratic in the GEM voltage to account for the large interval 
covered by the scans. 
The dashed curves are the predicted charge values for the upper layer 
and layer sum, according to the result of Fig.~\ref{fig:qcorr}, which again
illustrates the consistent behaviour of the charge sharing 
between the two readout layers.

The analysis of the cluster widths confirmed stable operating
conditions at all field values under study.

\subsection{Tracking Quality}

\noindent
To examine the tracking quality of the GEM detector, muon tracks were
reconstructed in the silicon telescope and extrapolated to the surface of 
the GEM detector. 
For this analysis, runs with different gains were used, taking into
account all clusters reconstructed in the upper and lower readout layers.
After alignment of the detectors, the GEM detector was
tested in two steps, 
\begin{enumerate}

\item First, the two readout layers were considered independently
      of each other, providing measurements of the coordinate 
      perpendicular to the pads of the muons traversing the detector.

\item Then combinations of the information of both layers were analysed.
      Of special interest are the sum and the difference of the
      position measurements. Owing to the small effective crossing angle 
      of the two layers, the position sum gives an improved determination 
      of the coordinate perpendicular to the pads.
      The difference in the positions provides a measure of the 
      coordinate along the pads of the readout structure.

\end{enumerate}

\subsubsection{Spatial Resolution}

\begin{figure}[htb]
\centering
\epsfig{file=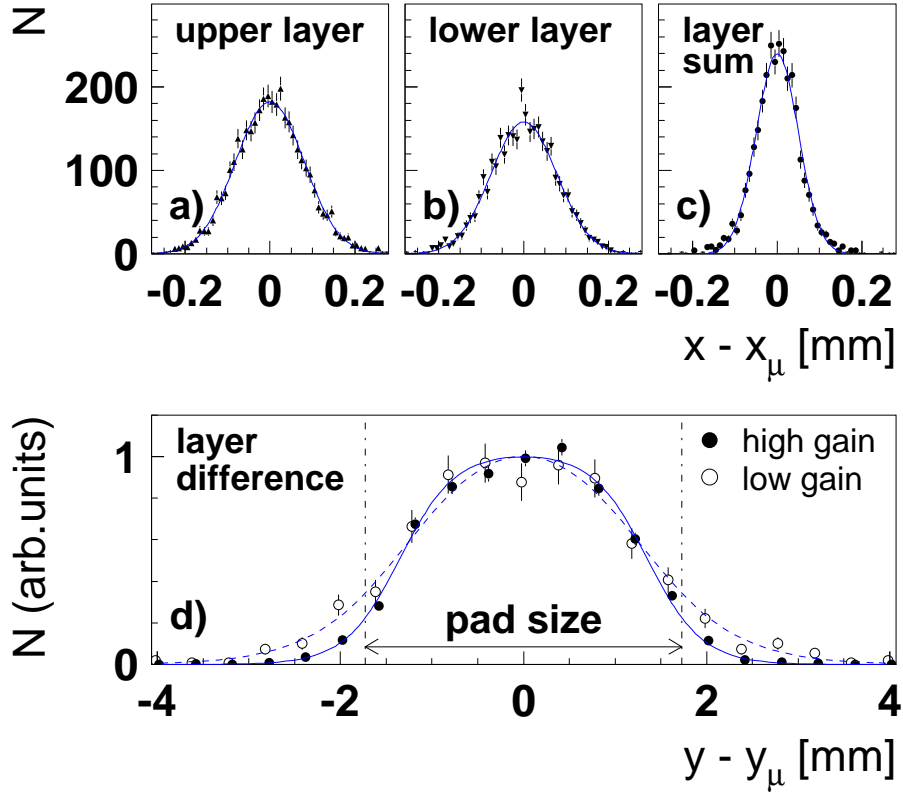,width=12.0cm}
\caption{
The position resolution for muons is shown for the 
a) upper layer, b) lower layer, c) layer sum, and d) layer difference.
The curves in a--c denote Gaussian fits to the data.
In d), measurements are shown at high/low gain
(solid/open circles)
together with fits to the data (curves, see text).}
\label{fig:resolution} 
\end{figure}

Examples of the differences between the predicted muon tracks and
the positions measured in the GEM detector are shown in 
Fig.~\ref{fig:resolution}.
The precision reached for the single layer measurements is found to be
$\sigma\approx80\,\mu$m as determined by the Gaussian fits, 
and is improved to $\sigma\approx50\,\mu$m when using both layers.
This resolution compares well with the resolution obtained 
in detectors with one-dimensional readout of 200\,$\mu$m pitch
\cite{mf2}.

The resolution of the determined coordinate along the pads exhibits a 
broad plateau, owing to the pad size of $3.4\,$mm being much larger
compared to the extent of the charge cloud.
Here two measurements are compared, where one was taken at high gain
(solid circles), and the other at low gain (open circles).
Both distributions have been fitted using a parameterization
with exponential edges, showing a full width at half maximum
of $2.8$\,mm and shorter tails for the measurement with the larger
signal clusters.

\begin{figure}[htb]
\setlength{\unitlength}{1cm}
\begin{picture}(15.0,10.0)
\put( 6.9,6.8){\rm\bf\large a)}
\put(11  ,6.8){\rm\bf\large b)}
\put(2.,0.0)
{\epsfig{file=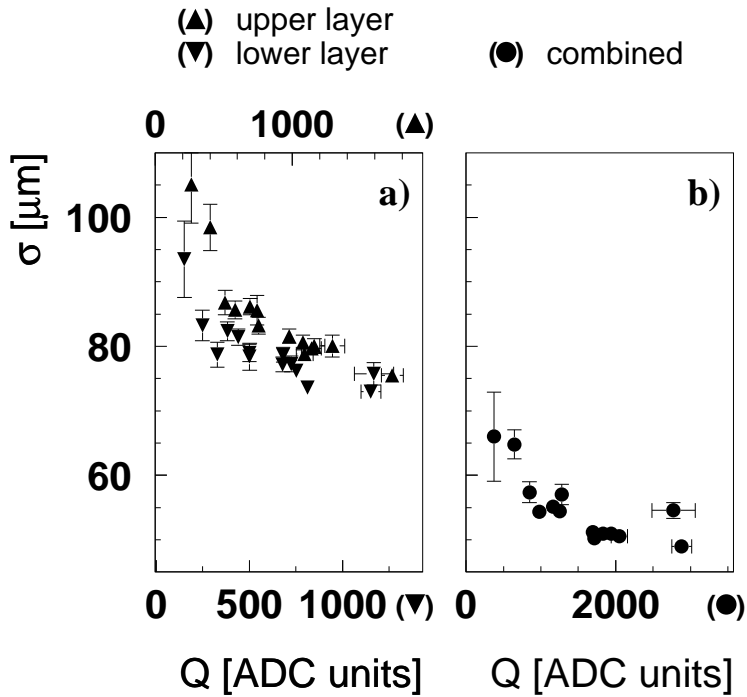,width=10.0cm}}
\end{picture}
\caption{
Position resolution of the coordinate perpendicular to the pads 
as a function of the relative gain $Q$. 
The upper charge scale is 
valid for the upper readout layer (up pointing triangle symbols) and the 
lower scale for the lower layer (down pointing triangles). The circles show 
the improved resolution when combining the position information from both 
readout layers.}  
\label{fig:resolution-scan} 
\end{figure}

In Fig.~\ref{fig:resolution-scan}, the dependence of the position
resolution $\sigma$ perpendicular to the pads
on the relative gain $Q$ of the clusters 
is shown, neglecting contributions from the finite resolution of the 
silicon telescope. 
The up/down pointing triangle symbols denote the upper/lower
single layer information, and the circles give the result
of the combined layer information.
As expected, the position measurements are best at high gain.

\begin{figure}[htb]
\centering
\epsfig{file=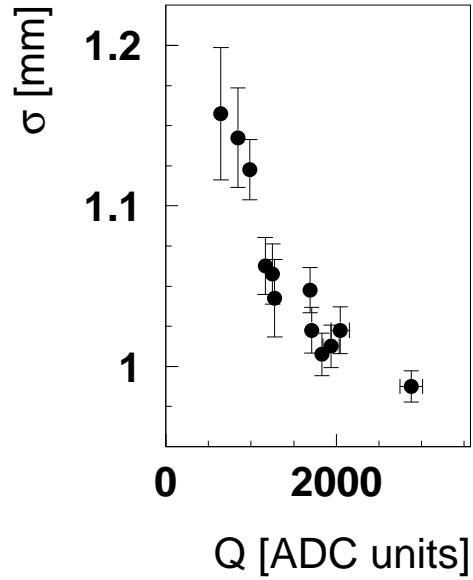,width=6.5cm}
\caption{
Position resolution of the coordinate measurement along the pads as 
a function of the summed most probable signal charge $Q$ (see text).}
\label{fig:sigmay} 
\end{figure}

The quality of the measurement of the coordinate along the $3.4\,$mm long 
pads is shown in Fig.~\ref{fig:sigmay} as a function of the summed 
most probable signal-charge of the two layers. 
To give values comparable to the Gaussian widths shown in 
Fig.~\ref{fig:resolution-scan}, we present here the width of the 
residual distribution containing 67\,\% of the detected events. 
We observe a resolution in the direction along the pads of 
$\sigma \approx 1$\,mm.

\subsubsection{Detection Efficiency and Rejection of Combinatorial Background}

\begin{figure}[htb]
\setlength{\unitlength}{1cm}
\begin{picture}(15.0,10.0)
\put( 6.9,2.4){\rm\bf\large a)}
\put(11  ,2.4){\rm\bf\large b)}
\put(2.,0.0)
{\epsfig{file=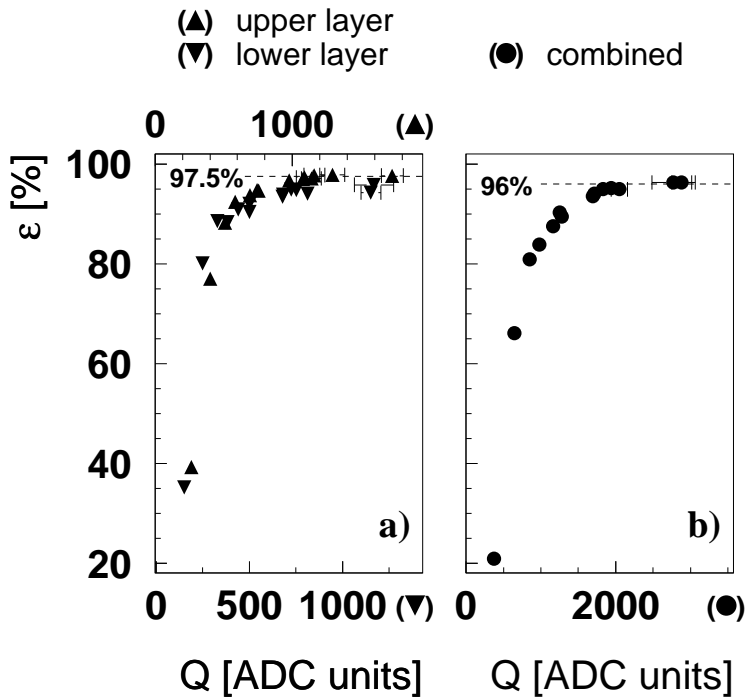,width=10.0cm}}
\end{picture}
\caption{
Muon detection efficiency as a function of the relative gain $Q$.
The upper charge scale is valid for the upper readout layer (up pointing 
triangle symbols) and the lower scale for the lower layer (down pointing 
triangles). The circles show the efficiency to detect a muon simultaneously 
in both readout layers.}  
\label{fig:efficiency} 
\end{figure}

The efficiency, equation (\ref{eq:efficiency}) in 
section~\ref{sec:anaTools}, of detecting a muon track in the single layers 
and their combinations is shown in Fig.~\ref{fig:efficiency} 
as a function of the relative gain $Q$.
The efficiencies reach $97.5\,\%/96\,\%$ for the upper/lower layer. 
The differences in efficiency can be attributed to the different 
fractions of the charge captured by in the two layers.  
Since both layer measurements are correlated due to the common
primary ionization process, the combined measurements reach
also efficiencies of $96\,\%$.

\begin{figure}[htb]
\setlength{\unitlength}{1cm}
\begin{picture}(15.0,10.0)
\put( 4.2,2.4){\rm\bf\large a)}
\put(8.35  ,2.4){\rm\bf\large b)}
\put(2.,0.0)
{\epsfig{file=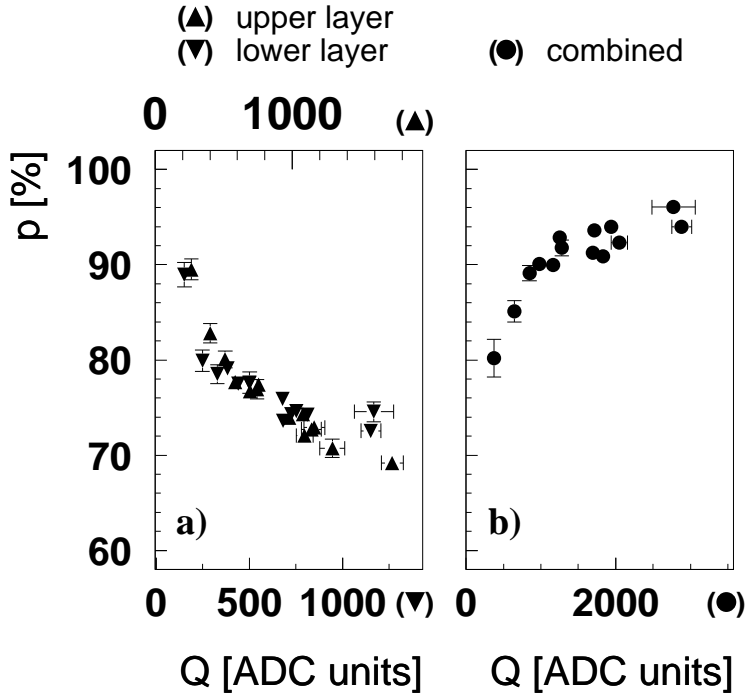,width=10.0cm}}
\end{picture}
\caption{
Purity of the cluster signals as a function of the relative gain $Q$. 
The upper charge scale is valid for the upper readout layer (up pointing 
triangle symbols) and the lower scale for the lower layer (down pointing 
triangles). The circles show the purity based on simultaneous 
detection in  both readout layers.}
\label{fig:purity} 
\end{figure}

To demonstrate the potential of rejecting ``ghost hits'' using the 
position and charge correlations between the two readout
layers, the purity equation~(\ref{eq:purity}) is studied in 
Fig.~\ref{fig:purity} as a function of the relative gain $Q$.
In Fig.~\ref{fig:purity}a, the purity of the single layer measurement 
(triangle symbols) is shown to decrease with increasing cluster charge to 
the level of $70\,\%$ in the region overlapping with the silicon telescope. 
This is understood to arise from the increasing number of channels 
exceeding the
noise threshold of the cluster definition.

Using the information of both layers 
provides an additional tool to reject the combinatorial ghost hits
(Fig.~\ref{fig:purity}b).
Restricting the combinations to the area of the telescope and exploiting 
the charge correlation shown in Fig.~\ref{fig:qcorr} 
rejects most wrong combinations and background 
hits with insignificant losses of the efficiency. 

\section{Conclusion}

\noindent
We have presented the construction of a large micro pattern gas detector 
with a new two-layer readout design, 
optimized for forward particle detection in the endcap of a collider 
experiment, 
and suited to low-cost mass production.
Using a high energy muon beam, we demonstrated that the detector 
is well operational under safe gas conditions over a wide range 
in the settings of the different electrical fields.
The small effective crossing angle of $6.7$ degrees of the two readout 
layers, with $406\,\mu$m pitch each, allows one coordinate to 
be measured with $\sigma\approx50\,\mu$m precision.
In addition, the perpendicular coordinate can be determined 
with a precision of $\sigma\approx1$\,mm at a pad length of $3.4\,$mm. 
The efficiency of detecting muons with the combined
layer measurements has been determined to reach $96\%$.

The performance tests give a significant part of the information 
required for applying the detector concept in a modern collider experiment.
Although our detector has not been tested in a high intensity
hadronic environment,
we remark that micro-pattern gas detectors with GEM foils have been 
operated successfully under LHC equivalent conditions \cite{mf2,bachmann}.

\section*{Acknowledgements}

\noindent
For interesting comments and discussions we 
wish to thank G.~Barker and S.~Kappler.
We thank our colleagues from the CERN accelerator devision for the
successful operation of the muon test beam.
For financial support we are grateful to the Bundesministerium f\"ur
Bildung und Forschung (BMBF 05-HC8VK17).
One of us (M.E.) wishes to thank Th.~M\"uller and the IEKP group of the 
University Karlsruhe for their hospitality, and the Deutsche 
For\-schungs\-ge\-mein\-schaft for the Heisenberg Fellowship.


\end{document}